\documentclass[prd, amsmath,amssymb,aps, reprint, preprintnumbers, nofootinbib, superscriptaddress]
              {revtex4-1}
              
\pdfoutput=1

\usepackage{amsmath}
\usepackage{amsfonts}
\usepackage{amssymb}
\usepackage{mathrsfs}
\usepackage{graphicx}
\usepackage{color}
\usepackage[dvipsnames]{xcolor}
\usepackage{longtable}
\usepackage{bm}
\usepackage{blindtext}
\usepackage{wasysym}
\usepackage{hyperref}
\hypersetup{colorlinks=true,allcolors=blue}
\usepackage[normalem]{ulem}

\newcommand{\bi}{\begin{itemize}}
\newcommand{\ei}{\end{itemize}}

\newcommand\Fontix{\fontsize{9}{11}\selectfont}

\begin{document}
 
\title{Probing dark matter inside Earth using atmospheric neutrino oscillations at INO-ICAL}
\author{Anuj Kumar Upadhyay}
\email{anuju@iopb.res.in (ORCID: 0000-0003-1957-2626)}
\affiliation{Department of Physics, Aligarh Muslim University, Aligarh 202002, India}
\affiliation{Institute of Physics, Sachivalaya Marg, Sainik School Post, Bhubaneswar 751005, India}
\author{Anil Kumar}
\email{anil.k@iopb.res.in (ORCID: 0000-0002-8367-8401)}
\affiliation{Institute of Physics, Sachivalaya Marg, Sainik School Post, Bhubaneswar 751005, India}
\affiliation{Applied Nuclear Physics Division, Saha Institute of Nuclear Physics, Bidhannagar, Kolkata 700064, India}
\affiliation{Homi Bhabha National Institute, Anushakti Nagar, Mumbai 400094, India}
\author{Sanjib Kumar Agarwalla}
\email{sanjib@iopb.res.in (ORCID: 0000-0002-9714-8866)}
\affiliation{Institute of Physics, Sachivalaya Marg, Sainik School Post, Bhubaneswar 751005, India}
\affiliation{Homi Bhabha National Institute, Anushakti Nagar, Mumbai 400094, India}
\affiliation{Department of Physics \& Wisconsin IceCube Particle Astrophysics Center, University of Wisconsin, Madison, WI 53706, U.S.A}
\author{Amol Dighe} 
\email{amol@theory.tifr.res.in (ORCID: 0000-0001-6639-0951)}
\affiliation{Tata Institute of Fundamental Research, Homi Bhabha Road, Colaba, Mumbai 400005, India}

\preprint{IP/BBSR/2021-12, TIFR/TH/21-22}

\date{\today}

\begin{abstract}

The interior of Earth's core can be explored using weak interactions of atmospheric neutrinos. This would complement gravitational and seismic measurements, paving the way for multimessenger tomography of Earth. Oscillations of atmospheric neutrinos passing through Earth are affected by the ambient electron density. We demonstrate that atmospheric neutrinos can probe the possible existence of dark matter inside Earth's core in a unique way --- by measuring the amount of baryonic matter using neutrino oscillations. We find that a detector like ICAL at INO with muon charge identification capability can be sensitive to dark matter with $\sim5\%-6\%$ mass of Earth, at 1$\sigma$ level with 500 kt yr exposure. We show that, while it will not be possible to identify the dark matter profile using neutrino oscillation experiments, the baryonic matter profile inside the core can be probed with atmospheric neutrinos.
  
\end{abstract}


\maketitle

\section{Introduction} 
What lies in the interior of Earth has been a longstanding puzzle and active research is being carried out in this direction. The regions deep below Earth's surface are inaccessible due to large temperatures, pressures, and extreme environments. Therefore, the information that can be obtained about them is only indirect, via gravitational~\cite{Luzum:2011,astro_alamanac} and seismic measurements~\cite{Robertson:1966}. Complementary to gravitational and seismic measurements, other probes can also be helpful viz. geoneutrino detection~\cite{Araki:2005qa,FIORENTINI2007117,McDonough2013,McDonough2015,Michael2017}, neutrino absorption~\cite{Gonzalez-Garcia:2007wfs,Donini:2018tsg}, and neutrino oscillations~\cite{Winter:2015zwx}. These complementary approaches have paved the way for the multimessenger tomography of Earth. Broadly speaking, Earth is composed of two concentric shells -- the outer one is mantle, and the inner one with a much higher density is the core. Some of the pressing issues regarding the details of the interior are (i) establishing the existence of a high-density core, (ii) measuring the location of the core-mantle boundary, (iii) determining the chemical composition of the core, and (iv) identifying the state of matter inside the core. These issues are crucial for a detailed understanding of the internal structure and dynamics of Earth.

The information about the internal structure of Earth, in terms of the radial distribution of its density, is obtained using the studies of propagation of seismic waves as they travel inside the Earth~\cite{Robertson:1966}. When seismic waves generated near the outer mantle travel through Earth, the varying density induces varying refractive index for the propagation of these waves and can bring some of them back to the surface due to total internal reflection. Wherever the density changes sharply, for example, at the core-mantle boundary, the waves also undergo partial reflection and refraction. Based on the data available on seismic waves, various models for Earth's density profile have been studied in the literature~\cite{Gilbert-Dziewonski1975, Dziewonski-Hales-Lapwood1975, Kennett-Engdahl1991, Kennett-etal1995, Cammarano-etal-2005, Kustowski-etal-2008}. Out of them, one of the most widely discussed model is the Preliminary Reference Earth Model (PREM)~\cite{Dziewonski:1981xy}.
 
The PREM profile is based on two empirical equations relating the velocities of the shear (S) and pressure (P) waves with the density of the layer they are passing through --- the Birch's law~\cite{Birch:1964} that is valid for the outer mantle, and the Adams-Williamson equation~\cite{Williamson:1923} that is valid for the inner mantle and core. Both of these are empirical relations, with parameters that depend upon temperature, pressure, composition, and elastic properties of Earth, which give rise to uncertainties. The uncertainty in the density of the mantle is about 5\%, whereas that for the core is significantly larger~\cite{Bolt:1991,kennett:1998,Masters:2003}.

The core is the least understood region of Earth. The outer core is inferred to be liquid because it has been observed that it does not allow the propagation of S waves, and the velocity of P waves drops sharply therein~\cite{Robertson:1966}. On the other hand, the inner core is likely to be solid since the velocity of the P waves is found to be higher in that region~\cite{Lehmann:1936,Brush:1980}. A recent study has claimed a decrease in the velocity of S waves in the inner core,  indicating the possibility of a soft inner core consisting of an exotic state of matter called superionic state~\cite{He:2022}. For an issue as important as the detailed internal structure of our planet, it is crucial to augment the seismological data with completely independent measurements. In this paper, we focus on obtaining information on the structure of the core, assuming that the total mass of the core is known and the density profile of the mantle is completely understood.

Neutrinos provide such an independent avenue for the determination of the internal structure of Earth, through their absorption and oscillations. The large cross sections of neutrinos at energies above a few TeV~\cite{Gandhi:1995tf} give rise to the attenuation of neutrinos passing through Earth. Using this to probe the internal structure of Earth has been proposed~\cite{Placci:1973,Volkova:1974xa}, and detailed studies involving neutrinos from different sources, such as man-made neutrinos~\cite{Placci:1973,Volkova:1974xa,Nedyalkov:1981,Nedyalkov:1981pp,Nedyalkov:1981yy,Nedialkov:1983,Krastev:1983,DeRujula:1983ya,Wilson:1983an,Askarian:1984xrv,Volkova:1985zc, Tsarev:1985yub,Borisov:1986sm,Tsarev:1986xg,Borisov:1989kh,Winter:2006vg}, extraterrestrial neutrinos~\cite{Wilson:1983an,Kuo:1995,Crawford:1995,Jain:1999kp,Reynoso:2004dt}, and atmospheric neutrinos~\cite{Gonzalez-Garcia:2007wfs,Borriello:2009ad,Takeuchi:2010,Romero:2011zzb}, have been carried out. This is often referred to as ``Earth tomography''~\cite{DeRujula:1983ya}. In Ref.~\cite{Donini:2018tsg}, the authors used the absorption of TeV-PeV neutrinos inside Earth, using one year of IceCube data, to determine its mass using weak interactions for the first time. Their results on the mass of Earth (with a 24\% error) are in agreement with present gravitational measurements. The exploration of Earth structure using diffraction patterns produced by coherent neutrino scattering inside Earth's matter does not seem to be technologically feasible~\cite{Fortes:2006}. 

The advancement in the precision of oscillation parameters with a nonzero value of reactor mixing angle $\theta_{13}$ has opened the door for Earth tomography based on matter effects on the oscillations of neutrinos in the multi-GeV energy range, which easily penetrate through the core. The possibility of such ``neutrino oscillation tomography'' has been studied using man-made neutrino beams~\cite{Ermilova:1986ph,Nicolaidis:1987fe,Ermilova:1988pw,Nicolaidis:1990jm,Ohlsson:2001ck,Ohlsson:2001fy,Winter:2005we,Minakata:2006am,Gandhi:2006gu,Tang:2011wn,Arguelles:2012nw}, solar neutrinos~\cite{Ioannisian:2002yj,Ioannisian:2004jk,Akhmedov:2005yt,Ioannisian:2015qwa,Ioannisian:2017chl,Ioannisian:2017dkx,Bakhti:2020tcj}, supernova neutrinos~\cite{Lindner:2002wm,Akhmedov:2005yt}, and atmospheric neutrinos~\cite{Agarwalla:2012uj,Rott:2015kwa,Winter:2015zwx,Bourret:2017tkw,Bourret:2019wme,Bourret:2020zwg,DOlivo:2020ssf,Kumar:2021faw,Maderer:2021aeb,Kelly:2021jfs,Denton:2021rgt,Capozzi:2021hkl,Upadhyay:2022jfd}. During this propagation, neutrinos undergo charged-current interactions with ambient electrons and experience an effective Earth matter potential 
\begin{equation}\label{eq:matter_pot}
\Fontix
V_\text{CC} = \pm\, \sqrt2 G_F N_e \approx \pm \,7.6 \times 10^{-14}\,
  Y_e\left(\frac{\rho}{\text{g/cm}^3}\right)\,\text{eV}\, ,
\end{equation}
where $Y_e = N_e/(N_p +N_n)$ is the relative number density of ambient electrons, and $\rho$ is the ambient matter density. The $\pm$ sign is for neutrinos and antineutrinos, respectively. Earth matter effects change the effective masses and mixing angles of neutrinos~\cite{Wolfenstein:1977ue,Mikheev:1986gs,Mikheev:1986wj}, and this change may be detected through its effect on the neutrino oscillation probabilities. These effects are prominent and distinct near neutrino energies of 5$-$10 GeV, where the Mikheyev-Smirnov-Wolfenstein resonance~\cite{Wolfenstein:1977ue,Mikheev:1986gs,Mikheev:1986wj} takes place. Moreover, when neutrinos with energies 2$-$6 GeV pass through the core of Earth, they experience the neutrino oscillation length resonance (NOLR)~\cite{Petcov:1998su,Chizhov:1998ug,Petcov:1998sg,Chizhov:1999az,Chizhov:1999he} or parametric resonance (PR)~\cite{Akhmedov:1998ui,Akhmedov:1998xq}.

Atmospheric neutrinos have energies in the multi-GeV range where Earth matter effects are significant. Since $\sim$ 20\% of the upward-going atmospheric neutrinos observed at a detector would have passed through the core, they would serve as probes of the core. The use of atmospheric neutrinos for constraining the average densities of the core and the mantle using ORCA~\cite{Winter:2015zwx,Maderer:2021aeb,Capozzi:2021hkl} and DUNE~\cite{Kelly:2021jfs}, for detecting the core-mantle boundary (CMB) using the iron calorimeter (ICAL)~\cite{Kumar:2021faw}, and for determining the position of the core-mantle boundary using DUNE~\cite{Denton:2021rgt} and ICAL~\cite{Upadhyay:2022jfd}, has already been proposed. Note that Earth matter effects on neutrinos depend on the electron density, as opposed to the overall matter density that the seismic waves are mainly sensitive to. Therefore, the information obtained from neutrinos is complementary to that obtained from seismology and can be used to gain information on the chemical composition of the core~\cite{Rott:2015kwa,Bourret:2017tkw,Bourret:2019wme,Bourret:2020zwg,Maderer:2021aeb}.

In this paper, we explore whether atmospheric neutrinos can also detect the presence of physics beyond the Standard Model (SM) of particle physics, at the same time probing the internal structure of the core. In particular, we analyze the effects of possible presence of dark matter (DM) in Earth on neutrino propagation, and whether these effects may be observed at an atmospheric neutrino detector. In the present study, we work in the context of a scenario where DM has minimal coupling with SM particles as well as with itself. Therefore, it does not directly affect neutrino oscillations. In other words, the effective Hamiltonian for neutrino propagation does not involve DM. However, the total amount of DM would matter because it determines how much baryonic matter is present inside Earth, and hence, how much electron number density is permitted. Therefore, the data could be sensitive to the total DM content inside Earth, though it will not be sensitive to the distribution of DM inside Earth. The presence of DM inside Earth may be indicated by the deviations of the neutrino oscillation probabilities from their expected values. Of course, other scenarios beyond the SM like nonstandard interactions can also cause such deviations, leading to possible degeneracies among new physics scenarios.  

Although there are strong constraints on the amount of ambient DM captured by Earth~\cite{Reno:2021cdh}, exotic scenarios like the formation of Earth around a DM seed may still be viable, similar to the formation of galaxies around DM cores~\cite{white:1978}, or the proposed scenarios of dark stars~\cite{Freese:2007cg} and primordial planets~\cite{Sivaram:2019}. Since DM in our scenario has minimal coupling with SM particles (just enough to allow the formation of Earth), the cross section for direct interaction of DM with SM, as well as the cross section for annihilation of DM to SM particles,  will be extremely small. Therefore, no signal is expected at the direct or indirect DM search experiments, and  this scenario will not be in conflict with any of the current bounds~\cite{Arbey:2021gdg} on the relevant cross sections.

In any case, it is always desirable to obtain constraints from completely independent means, and further, the capabilities of new detectors for obtaining such constraints are worth examining. In that spirit, in this paper, we would not like to go deeper into the question of the extent of DM inside Earth. Our aim is to determine the sensitivity of an atmospheric neutrino experiment like the proposed ICAL detector at the India-based Neutrino Observatory (INO)~\cite{ICAL:2015stm} to the presence of DM inside Earth. Note that the results of Ref.~\cite{Donini:2018tsg} using one-year IceCube data may be interpreted as ruling out the presence of more than 32\% (24\%)  of the mass of the core (Earth) in the form of DM at $1\sigma$ level. The mass of Earth has also been estimated using matter effects in atmospheric neutrino oscillations at the Super-K experiment~\cite{Super-Kamiokande:2017yvm}. Their results can be interpreted as ruling out more than 21\% mass of the entire Earth in the form of DM at $1\sigma$ level. There are also a few more sensitivity studies for measuring the mass of Earth which can be interpreted as constraining the amount of DM inside Earth. For example, the future atmospheric experiment like the ARCA detector may be expected to constrain DM up to 25\% of the mass of Earth at $1\sigma$ in ten years using neutrino absorption~\cite{Maderer:2022obm}. The ORCA detector may be expected to constrain DM to be less than about 6\% of the mass of Earth at $1\sigma$ in the same time, using neutrino oscillations~\cite{Maderer:2022obm}. Similarly, atmospheric DUNE experiment may be expected to constrain DM to be less than about 9\% of the mass of Earth at $1\sigma$ in ten years~\cite{Kelly:2021jfs}.

\begin{figure}
	\centering
	\includegraphics[width=\linewidth]{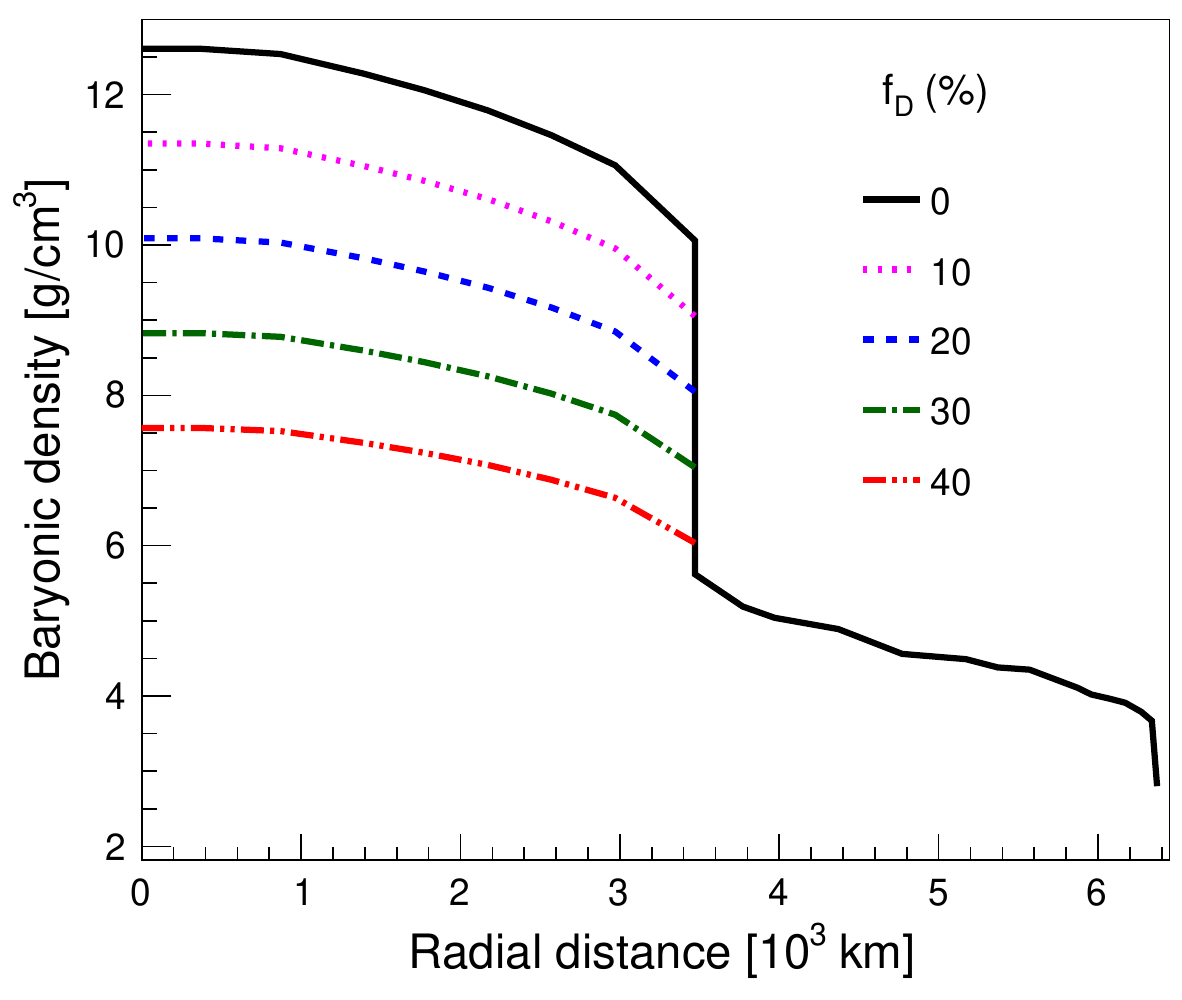}
	\caption{Representative  baryonic density profiles of Earth, obtained with a uniform $f_\text{B}(r) =  \langle f_\text{B} \rangle$ inside the core. The curves are labeled by the parameter $f_\text{D} \equiv 1 - \langle f_\text{B} \rangle$. In the absence of DM, the density is taken to be the 25-layered PREM profile.}
	\label{fig:DM_fraction}
\end{figure} 

\begin{figure*}
	\centering
	\includegraphics[width=0.49\linewidth]{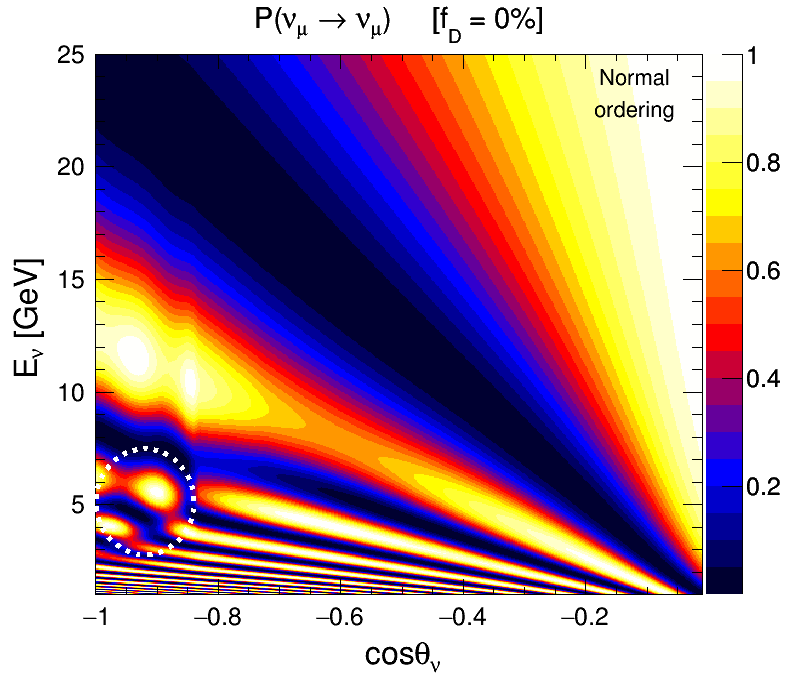}
	\includegraphics[width=0.49\linewidth]{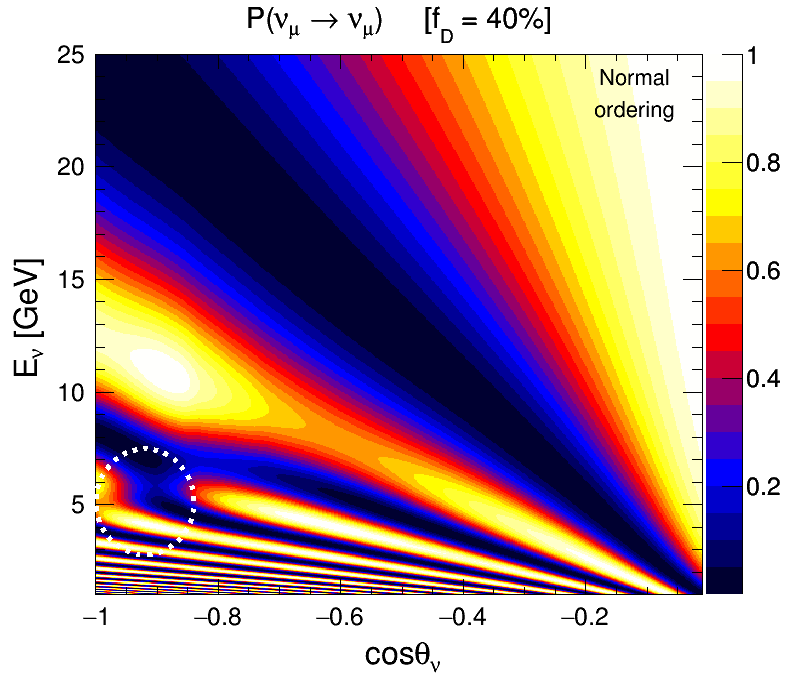}
	\caption{Three-flavor neutrino survival probability for $\nu_\mu \rightarrow \nu_\mu$ in Earth in the ($E_\nu$, $\cos\theta_\nu$) plane. Left: PREM profile   without dark matter. Right: modified PREM profile with $f_\text{D}=40\%$ in the core (see the dash-dot-dotted red curve showing the corresponding baryonic density profile in Fig.~\ref{fig:DM_fraction}). The NOLR/PR around (5~GeV, -0.9) is highly diluted in the right panel. }
	\label{fig:puu-nu-mat25}
\end{figure*}

The mass of Earth has been determined to a very high precision using gravitational measurements~\cite{Luzum:2011,astro_alamanac}. These measurements are sensitive to the sum of the DM and baryonic mass, and hence would not be able to distinguish between the two. We denote the baryonic mass fraction of the core as 
\begin{equation}
\rho_\text{B}(r) = f_\text{B}(r)\, \rho_\text{PREM} (r)\,,
\end{equation}
and the average baryonic mass fraction of the core as
\begin{equation}\label{eq:f_B}
\langle f_\text{B} \rangle \cdot M_{\rm core} =
\int_0^{R_\text{CMB}} 4 \pi r^2  \rho_\text{B}(r)dr \; ,
\end{equation}
where $\rho_\text{PREM}(r)$ and $\rho_\text{B}(r)$ are the density distributions for the standard PREM profile and the actual baryonic profile, respectively, and $R_\text{CMB}$ is the core-mantle boundary. Since the difference between $M_\text{core}$ and $\langle f_\text{B} \rangle \cdot M_{\rm core}$ should be compensated by the DM, we use the parameter $f_\text{D} \equiv 1 - \langle f_\text{B} \rangle$ to quantify the amount of DM inside Earth. The correspondence between the parameter $f_\text{D}$ and the fraction of DM inside Earth is given in Table~\ref{tab:DM_fractions} for some representative values. Note that, since neutrino oscillations depend only on the electron number density and are blind to the DM, the neutrino oscillation experiments will be sensitive only to the value of $f_\text{D}$ and not to the specific DM profile that gives rise to that value of $f_\text{D}$.

\begin{table}
	\centering
	\begin{tabular}{|c|c|c|}
		\hline \hline
		$f_\text{B} (\%) $ & $f_\text{D} (\%)$  & $M_\text{DM}/M_\text{E} (\%)$ \\
		\hline
	 	100	& 0 & 0 \\
		\hline
		90 & 10 & 3.2 \\
		\hline
		80 & 20 & 6.5 \\
		\hline
		70 & 30 & 9.7 \\
		\hline
		60 & 40 & 13.0 \\
		\hline \hline
	\end{tabular}
	\label{tab:DM_fractions}
	\caption{Some representative values of DM fractions. Here, $M_\text{E}$ is the mass of Earth and $M_\text{DM}$ is the mass of DM inside Earth. }
\end{table}

To start with, we choose a uniform $f_\text{B}(r)$. That is, the mass of the DM is compensated by decreasing the density of the core by a uniform fraction $f_\text{D}$. We further require that, at the core-mantle boundary, the baryonic density of the core is greater than that of the mantle, i.e. $\rho_\text{B}(R_{\rm CMB}^-) > \rho_\text{B}(R_{\rm CMB}^+)$. This implies $0 \leq f_\text{D} \leq 0.4$. Figure~\ref{fig:DM_fraction} shows the baryonic density profiles based on 25-layered PREM profile, with some representative values of $f_\text{D}$.

\section{Oscillation probabilities with atmospheric neutrinos}

Atmospheric neutrinos cover a wide range of energies (multi-GeV) and  baselines ($\sim15$--$12750$ km). Our numerical results have been presented by using the benchmark values of oscillation parameters as $\sin^2 2\theta_{12} = 0.855$, $\sin^2 \theta_{23} = 0.5$, $\sin^2 2\theta_{13} = 0.0875$, $|\Delta m^2_{32}| =2.46 \times 10^{-3}$ eV$^2$, $\Delta m^2_{21} =7.4 \times 10^{-5}$ eV$^2$, and $\delta_\text{CP} = 0$. These values are consistent with current global fits of neutrino data~\cite{NuFIT,deSalas:2020pgw,Capozzi:2021fjo}. We keep the values of these parameters fixed, since at the timescales of 20 years when our results on Earth tomography will be significant, these parameters are expected to be measured with a high precision~\cite{Song:2020nfh}. For the neutrino mass ordering, which is also expected to be determined over this timescale, we consider normal ordering for our analysis. We illustrate our results for the isoscalar Earth, where $N_n = N_p = N_e$, or $Y_e = 0.5$.

Figure~\ref{fig:puu-nu-mat25} shows the three-flavor oscillograms for survival probabilities of $\nu_\mu$ with energy $E_\nu$ and arrival zenith angle $\theta_\nu$, for the PREM profile without dark matter (left panel) and with 40\% dark matter fraction in the core (right panel). The difference in the oscillation probabilities is apparent, especially in the  NOLR/parametric resonance region near the left bottom region.

\section{Earth tomography with atmospheric neutrinos at INO-ICAL}
As an example of an atmospheric neutrino detector, we consider the proposed ICAL detector at INO~\cite{ICAL:2015stm}. ICAL is sensitive mainly to muons produced from charged-current interactions of atmospheric muon neutrinos and antineutrinos at multi-GeV energies and can identify the muon charge. The 50 kton ICAL detector has an energy resolution of 10\%--15\% for muons in the energy range of 1--25 GeV and an excellent angular resolution of better than 1$^\circ$. The magnetic field of about 1.5 Tesla~\cite{Behera:2014zca} enables ICAL to distinguish between $\mu^-$ and $\mu^+$ events and hence, between muon neutrinos and antineutrinos. We will show later that this charge identification (CID) capability of ICAL would play an important role in measuring the amount of DM inside Earth. Other atmospheric neutrino experiments such as IceCube/DeepCore~\cite{Gonzalez-Garcia:2007wfs,Donini:2018tsg}, Super-K/Hyper-K~\cite{Hyper-Kamiokande:2018ofw}, DUNE \cite{Kelly:2021jfs,Denton:2021rgt}, and ORCA \cite{Capozzi:2021hkl} can also perform this study. Though these detectors may not have CID capability, they can take advantage of their sensitivity to electron neutrinos.

We simulate the unoscillated events at ICAL using the NUANCE~\cite{Casper:2002sd} neutrino event generator with the Honda 3D flux at the INO site~\cite{Athar:2012it,Honda:2015fha}. We incorporate the full three-flavor oscillation probabilities in the presence of Earth matter effects using the reweighting algorithm~\cite{Ghosh:2012px,Thakore:2013xqa}. The detector responses for muons~\cite{Chatterjee:2014vta} and hadrons~\cite{Devi:2013wxa} have been incorporated using the ICAL  migration matrices. To calculate the expected median sensitivity of ICAL to the presence of DM, we use the reconstructed muon energy $E_\mu^\text{rec}$, muon direction $\cos\theta_\mu^\text{rec}$, and hadron energy ${E'}_\text{had}^\text{rec} \equiv (E_\nu - E_\mu)^\text{rec}$ as observables~\cite{Devi:2014yaa}, the binning scheme optimized for matter effects in Earth core~\cite{Upadhyay:2022jfd}, and the procedure for calculating $\Delta \chi^2$ outlined in~\cite{Devi:2014yaa}. The results are quantified in terms of $\Delta\chi^2_\text{DM}$ between the model with DM and the scenario without DM for which data are simulated.

\section{Can the DM mass fraction inside the core be measured?}
\begin{figure}
  \centering
  \includegraphics[width=\linewidth]{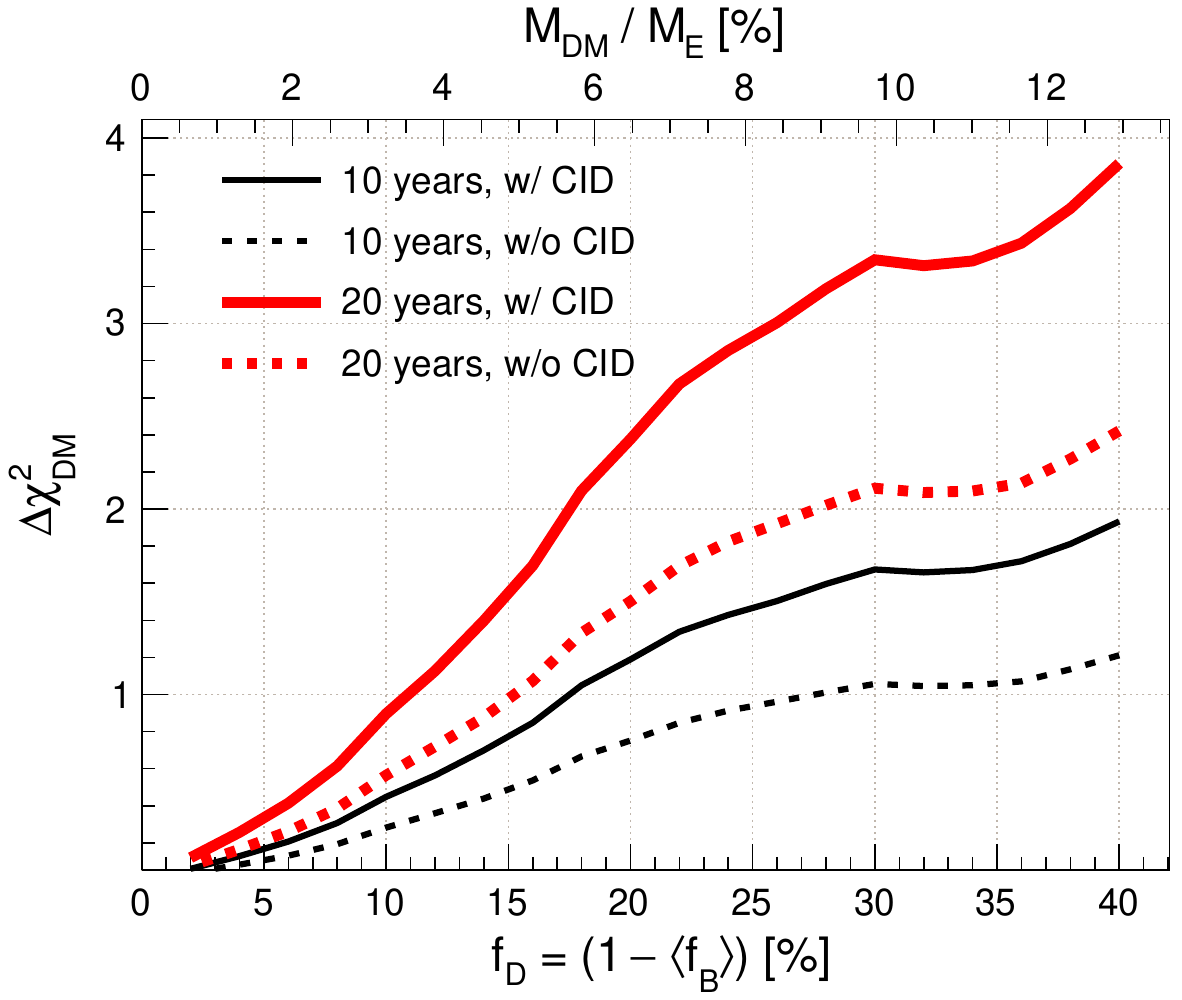}
  \caption{The sensitivity in terms of $\Delta\chi^2_\text{DM}$ with which ICAL can rule out a given value of $f_\text{D}$, for a uniform $f_\text{B}(r) = 1 - f_\text{D}$. The scale on the top of the figure indicates the corresponding fraction of DM mass inside Earth. The results are shown for 10 (thin curves) and 20 (thick curves) years of exposure with (solid curves) and without (dashed curves) CID.}
  \label{fig:DM_inside_Earth}
\end{figure} 
Figure~\ref{fig:DM_inside_Earth} presents the sensitivity at which $f_\text{D}$ may be excluded, in terms of the $\Delta\chi^2_\text{DM}$ value corresponding to it, compared with the hypothesis of no DM. The results are presented for the PREM profile with 25 layers. The Figure shows that the sensitivity to DM increases with $f_\text{D}$: for $f_\text{D} \sim 0.4$, it reaches $\Delta \chi^2_\text{DM} \approx 4$ (i.e. 2$\sigma$) with an exposure of 20 years, utilizing the CID capability of ICAL.  Without this CID capability, the sensitivity for DM would be lower by almost 40\%. This is because the matter effects modify $\nu$ and $\bar{\nu}$ oscillation probabilities differently, and CID capability helps to preserve this information. Note that, since the mass of the core is apporoximately 30\% of the Earth mass, $f_\text{D} \sim 0.4$ corresponds to DM comprising of about 12\% of Earth mass.

Figure~\ref{fig:DM_inside_Earth} shows that ICAL may be expected to be sensitive to DM at $1\sigma$ level in 10 (20) years if the DM forms $\sim5.5\%$ ($3.5\%$) of the mass of Earth. Currently running neutrino experiments like IceCube and Super-K would have collected a large amount of events by the time ICAL comes online and may have reached better than $\sim10\%$ sensitivity at $1\sigma$ with improved analysis techniques. However, once ICAL starts taking data, within a decade it would become quite competitive with the reach of other experiments, including the future ones that have been mentioned in the introduction. This is not unexpected, since ICAL is optimized for the multi-GeV range of energies where the matter effects, crucial for DM detection as indicated in this paper, are significant. The ability of ICAL to separate neutrino and antineutrino modes allows us to measure these matter effects efficiently.

It must be emphasized that the neutrino oscillation experiments have sensitivity to $f_\text{D}$, but not to the DM density profile, since neutrinos interact only with baryonic matter. However, realistic DM density profiles can be accommodated for a given $f_\text{D}$ as discussed in Appendix~\ref{sec:DM_distribution}.

\section{Identifying the baryonic matter profile inside the core}
\begin{figure*}
	\centering 
	\includegraphics[width=0.45\linewidth]{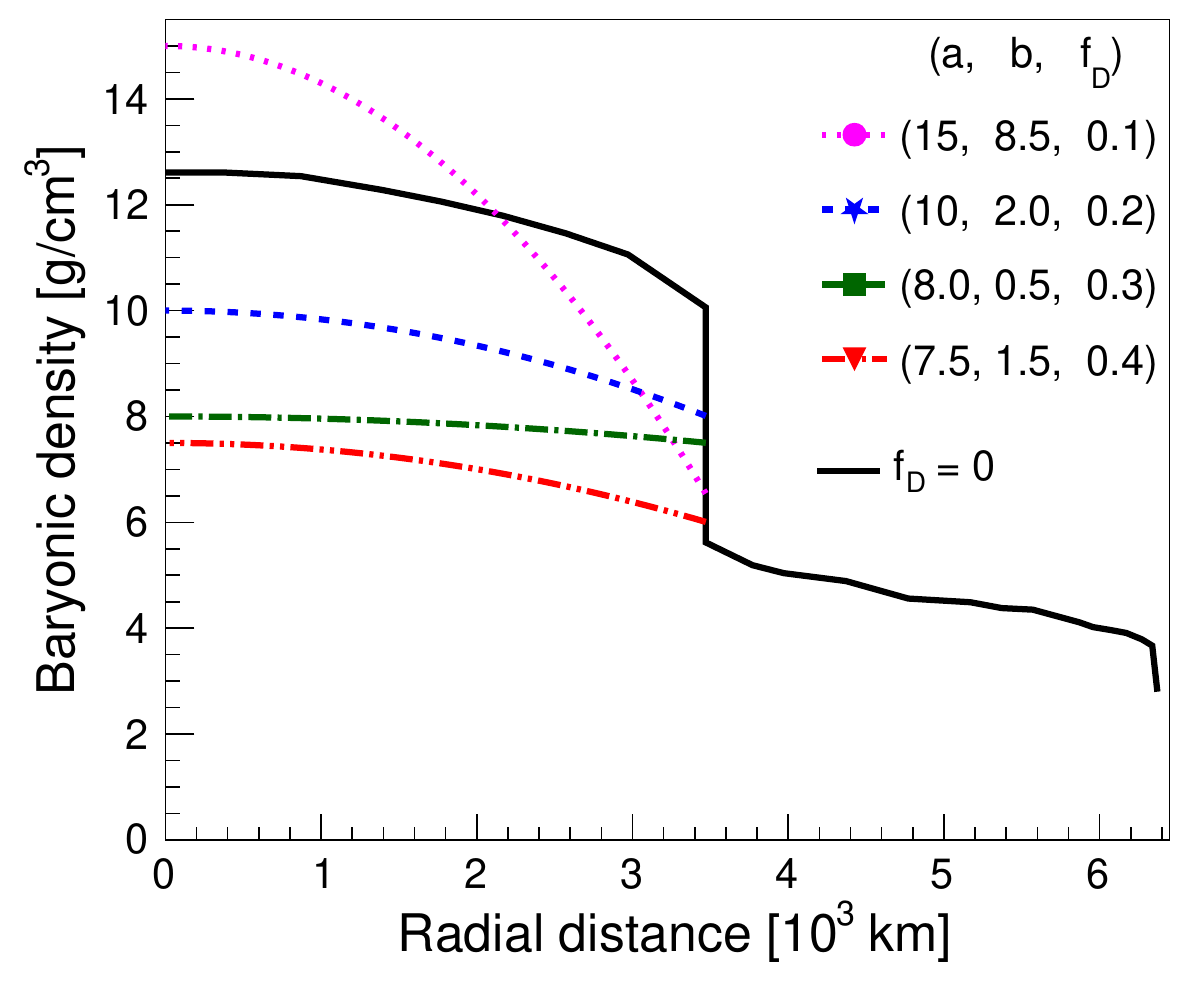}
	\includegraphics[width=0.45\linewidth]{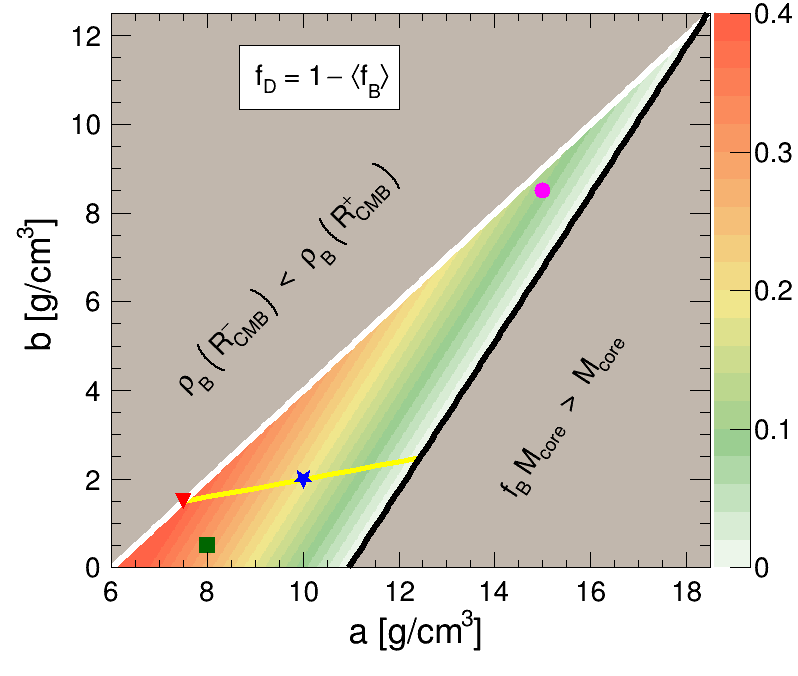}
	\includegraphics[width=0.45\linewidth]{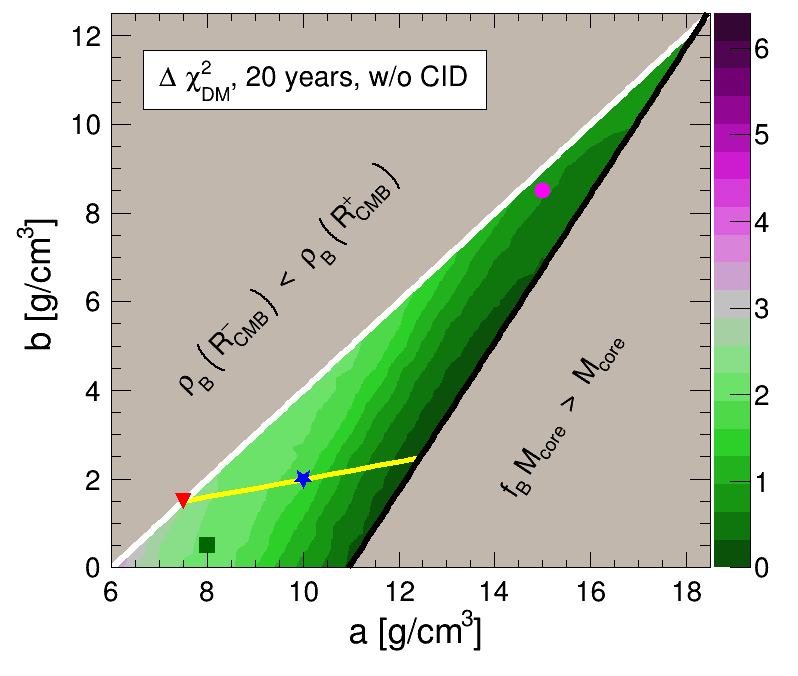}
	\includegraphics[width=0.45\linewidth]{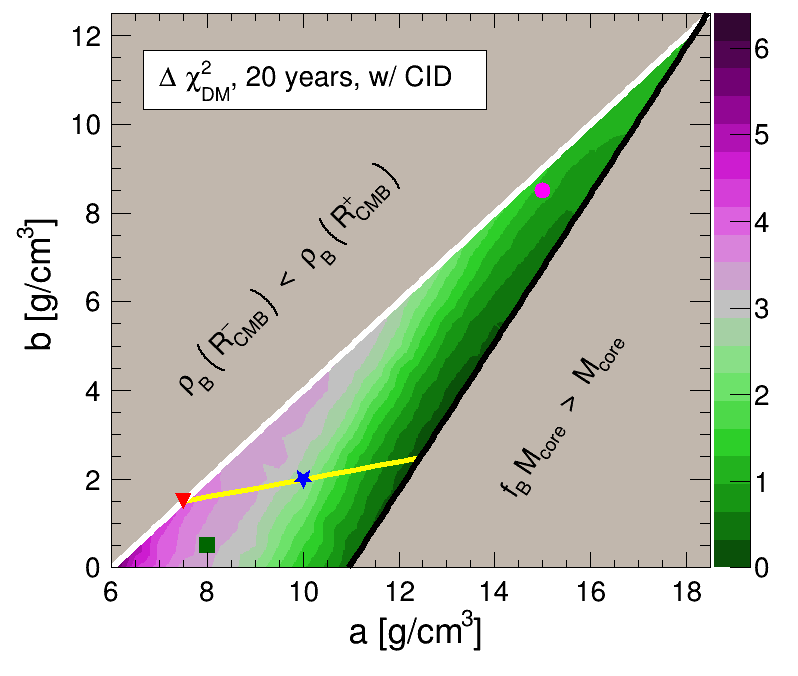}
	\caption{Top left: some representative baryonic density profiles with the form $\rho_B(r) = a - b (r / R_{\rm CMB})^2$ and their DM fraction $f_\text{D}$. Top right: color gradients of $f_\text{D}$ inside the core in the $(a,b)$ plane, such that $0 \leq f_\text{D} \leq 0.4$. The gray regions are unphysical. Bottom: color gradients of $\Delta \chi^2_\text{DM}$ in the $(a,b)$ plane, (left) without CID and (right) with CID. The markers with different colors and types showing density profiles in the top left correspond to the markers with the same colors and types in the other three panels. The yellow line of the form $b = \gamma a$ represents the various core density profiles (as shown in Fig.~\ref{fig:DM_fraction}) that look almost similar to the scaled-down versions of the 25-layered PREM profile inside the core.}
	\label{fig:Baryonic_profile}
\end{figure*}

Though neutrino oscillations cannot identify the DM density profile inside Earth's core, they can constrain the baryonic density profile inside the core. In order to illustrate this, we parameterize the baryonic matter density profile inside the core as
\begin{equation}
  \rho_B(r) = a - b \cdot \left(r/R_{\rm CMB} \right)^2 \; ,
  \label{eq:ab}
\end{equation}
where $a$ and $b$ are positive constants with the units of density. This profile has the desired property of monotonic decrease of density with radius. We ensure $\rho_\text{B}(R_\text{CMB}^{-}) > \rho_\text{B}(R_\text{CMB}^{+})$. Further restricting
\begin{equation}
  \int_0^{R_{\rm CMB}} 4 \pi r^2 \rho_B(r) dr
  \leq  M_{\rm core} \; 
\end{equation}
makes sure that the baryonic mass inside the core is less than the total mass of the core. In Fig.~\ref{fig:Baryonic_profile}, each point in the $(a,b)$ plane inside the triangular region formed by the white line, the black line, and the x axis corresponds to an allowed baryonic profile. We already know that the remaining mass may always be attributed to the DM. 

The top right panel of Fig.~\ref{fig:Baryonic_profile} shows the contours of $f_\text{D}$ within the core. The points along the black line correspond to the profiles without DM, i.e. $f_\text{D}=0$. Moving along a yellow line of the form $b = \gamma a$ corresponds to the scaling of a baryonic profile, with the value of $f_\text{D}$ changing along the line segment, which represents the core density profiles (like those in Fig.~\ref{fig:DM_fraction}) that look almost similar to the scaled-down versions of the 25-layered PREM profile inside the core.
 
The bottom panels in Fig.~\ref{fig:Baryonic_profile} show the sensitivity of the 50 kton ICAL to the baryonic matter profile, using 20 years of exposure. It is clear that the CID capability is crucial for an enhanced sensitivity for DM in Earth's core.  Note that the contours of $\Delta \chi^2_\text{DM}$ in the lower panels are close to, but not identical with, the contours of $f_\text{D}$ in the top right panel. This indicates that, while $f_\text{D}$ is the major factor determining the sensitivity of neutrino data to DM, it is not the only one; the neutrino data is sensitive to both the parameters $a$ and $b$. Indeed, even in the complete absence of DM ($f_\text{D} = 0$), moving along the black line  to higher values of $a$ and $b$ can be seen to change the sensitivity, albeit to a small extent. Therefore, these results are applicable even in the scenario without DM and indicate the sensitivity of ICAL to different density profiles of the core. 

\begin{figure*}
	\centering 
	\includegraphics[width=0.45\linewidth]{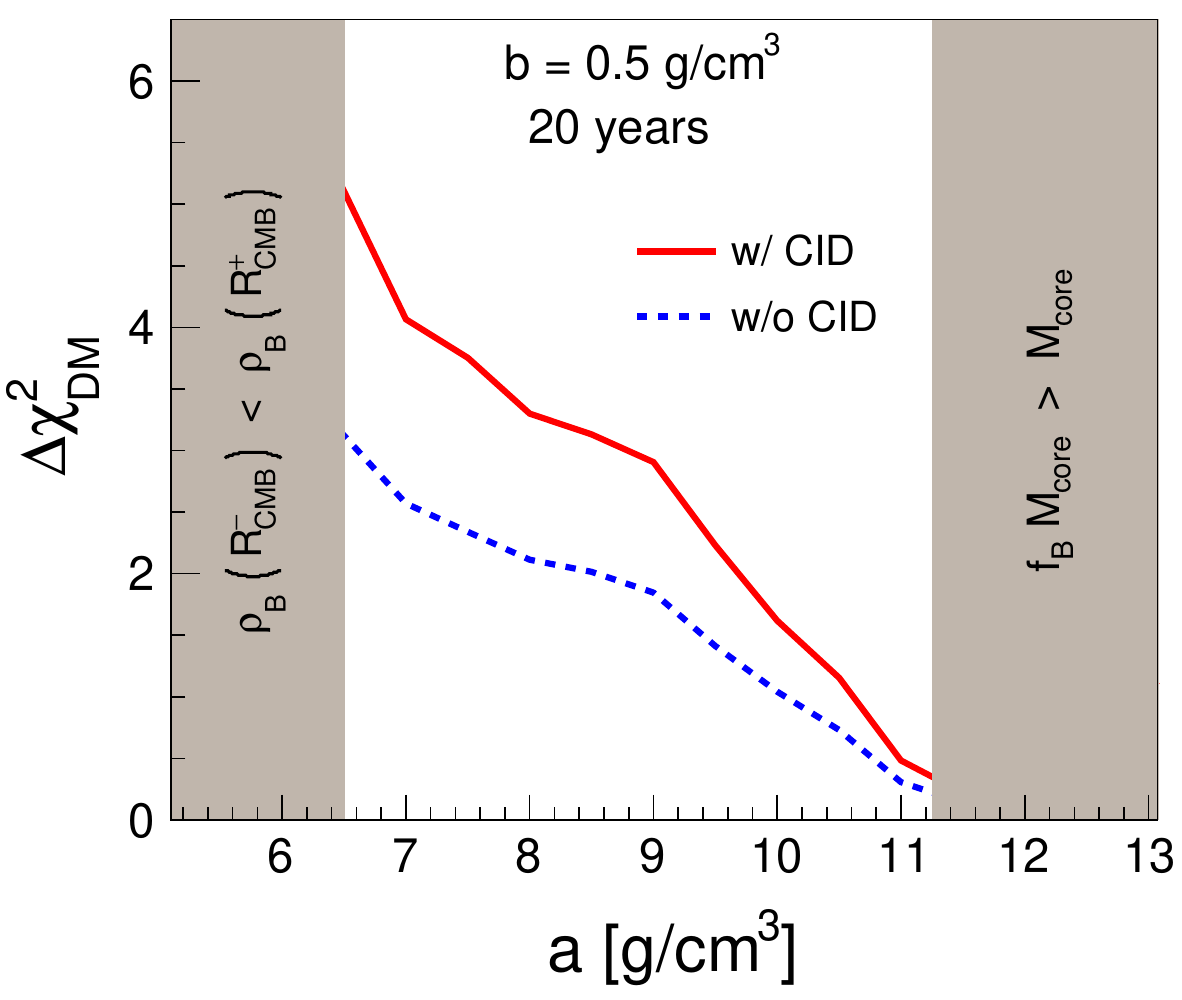}
	\includegraphics[width=0.45\linewidth]{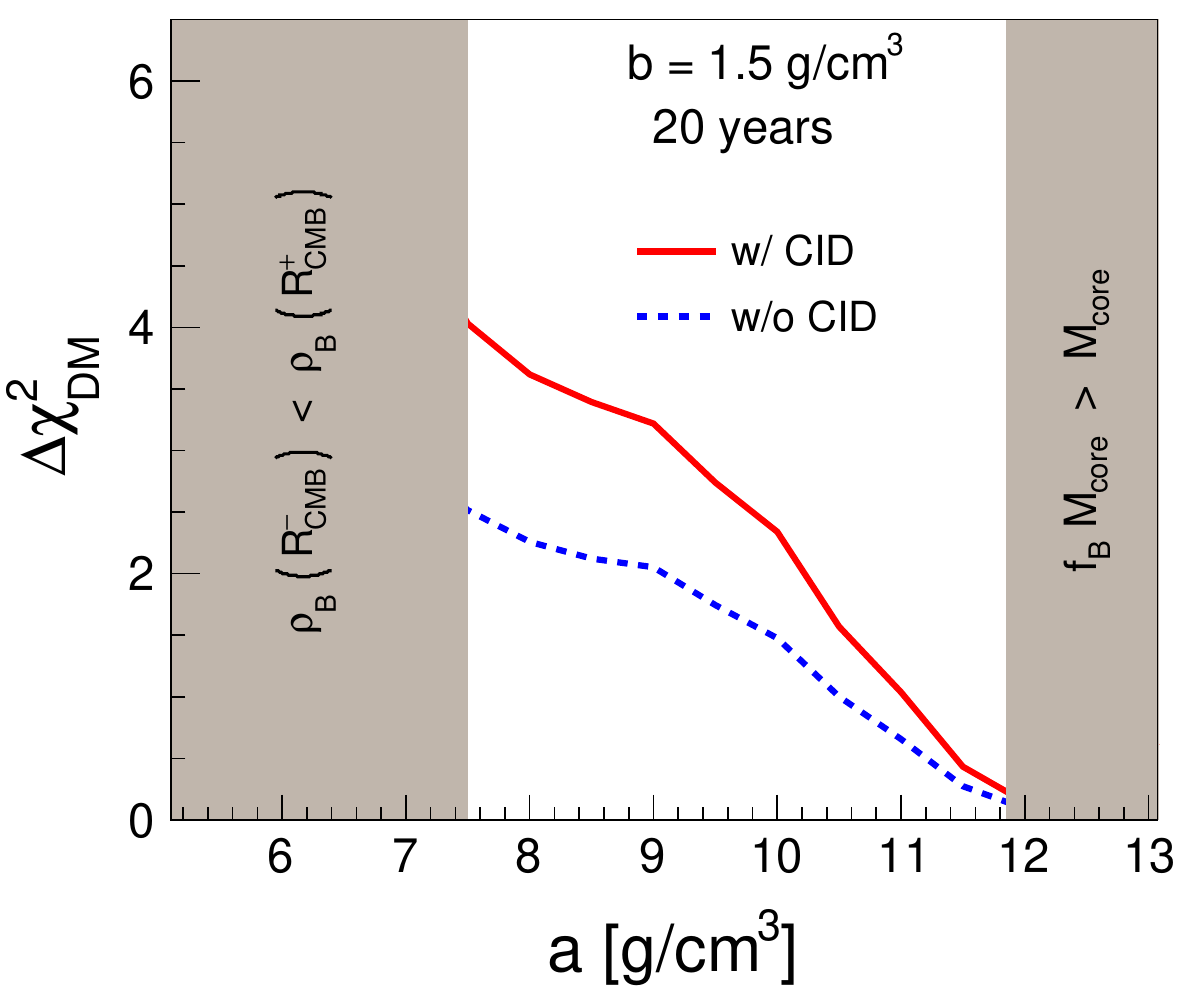}
	\caption{The value of $\Delta \chi^2_\text{DM}$ at ICAL as functions of parameter $a$ for two benchmark choices of parameter $b$, viz. $b = 0.5$ (left) and $b = 1.5$ (right), respectively. The solid red (dashed blue) curve shows the sensitivity with (without) the CID capability. The shaded regions at low and high values of $a$ correspond to those excluded by physics consideration as in Fig.~\ref{fig:Baryonic_profile}.}
	\label{fig:Baryonic_profile_projection}
\end{figure*}

We also present our results in another form in Fig.~\ref{fig:Baryonic_profile_projection}, where $\Delta \chi^2_\text{DM}$ is shown as function of $a$ for two benchmark choices of $b = 0.5$ (left panel) and $1.5$ (right panel). These choices are motivated by the observations that $b = 0.5$ is close to the region where we have maximum sensitivity and 1.5 is close to the yellow line shown in Fig.~\ref{fig:Baryonic_profile}. We can observe that $\Delta \chi^2_\text{DM}$ is larger at smaller value of $a$, i.e., data can more efficiently rule out those scenarios where the maximum density in core is lower. ICAL can disfavor the scenario with $(a,b) = (6.5,0.5)$ with   $\Delta \chi^2_\text{DM} \approx 5$ using 20 years of data.

\section{Concluding remarks}
The internal structure of Earth has been explored with gravitational interactions and using seismic waves, which are essentially electromagnetic interactions. In this paper, we point out a way to use weak interactions to probe the structure of Earth's core and show that this is feasible at upcoming neutrino experiments.

We focus on detecting the presence of DM inside the core of Earth. In such a scenario, the three kinds of measurements above are sensitive to three different quantities. While gravitational experiments do not distinguish between normal matter and DM and hence are sensitive to their sum, seismological studies are sensitive to only the normal matter, i.e., to the atoms and molecules. Neutrinos are unique in their sensitivity only to the electron number density inside Earth. These three measurements are thus complementary to each other.

Atmospheric neutrinos have substantial flux in the energy range of 1-10 GeV, where the Earth matter effects on the oscillations of core-passing neutrinos are significant. We show that a detector like ICAL, which is sensitive to multi-GeV neutrinos and can identify whether they are neutrinos or antineutrinos, would be able to rule out certain density profiles if the baryonic mass inside the core turns out to be smaller than expected mass of the core. This missing mass can be attributed to the dark matter. The sensitivity to a particular baryonic profile depends mainly, but not entirely, on the net DM fraction. The neutrino data are insensitive to the DM density profile.

We show that, within a decade after it starts operation, the reach of ICAL in establishing DM inside Earth would be comparable to what we expect from the presently running and proposed experiments.
Moreover, ICAL will enable us to test the existence of dark matter inside Earth in neutrino and antineutrino modes separately. Even in the absence of DM, the ICAL data will be sensitivity to the baryonic density profile.

Our results suggest that the exclusion of tens of percent of DM mass fractions inside a core to more than $3\sigma$ is expected to take decades. However, for a goal as crucial as having a better understanding of the internal structure of the planet we all live on, it is worthwhile to have a coordinated effort over a long time. Indeed, most of the planned neutrino detectors will keep on taking data for long periods. Such detectors will also play a major role in multi-messenger astronomy, since they will be sensitive to astrophysical phenomena that emit neutrinos. A detector like ICAL, capable of identifying the muon charge, will have a special advantage, as described in this paper.

\begin{acknowledgments}
This work is performed by members of the INO-ICAL collaboration to explore the possibility of utilizing oscillations of atmospheric neutrinos deep inside Earth to look for dark matter inside the core. We thank F. Halzen, A. Donini, S. Palomares-Ruiz, and R. Gandhi for their useful comments on our work. We sincerely thank A. Raychaudhuri and S. Goswami for  their careful reading of the manuscript and for providing useful suggestions. We acknowledge the support of  the Department of Atomic Energy (DAE), Govt. of India, under the Project Identification Numbers RTI4002 and RIO 4001. S.K.A. is supported by the Young  Scientist Project [INSA/SP/YSP/144/2017/1578] from the Indian National Science Academy (INSA). S.K.A.  acknowledges the financial support from the Swarnajayanti Fellowship Research Grant (No. DST/SJF/PSA- 05/2019-20) provided by the Department of Science and  Technology (DST), Govt. of India, and the Research Grant (File no. SB/SJF/2020-21/21) provided by the  Science and Engineering Research Board (SERB) under the Swarnajayanti Fellowship by the DST, Govt. of India. S.K.A would like to thank the United States-India Educational Foundation for providing the financial support from the the Fulbright-Nehru Academic and Professional Excellence Fellowship (Award No. 2710/F-N APE/2021). A.K.U. acknowledges financial support from the DST, Govt. of India (DST/INSPIRE Fellowship/2019/IF190755). The numerical simulations are performed using the SAMKHYA: High-Performance Computing Facility at the Institute of Physics, Bhubaneswar.
\end{acknowledgments}

\appendix

\section{Accommodating a realistic DM profile }
	\label{sec:DM_distribution}
\begin{figure*}
	\centering 
	\includegraphics[width=0.5\linewidth]{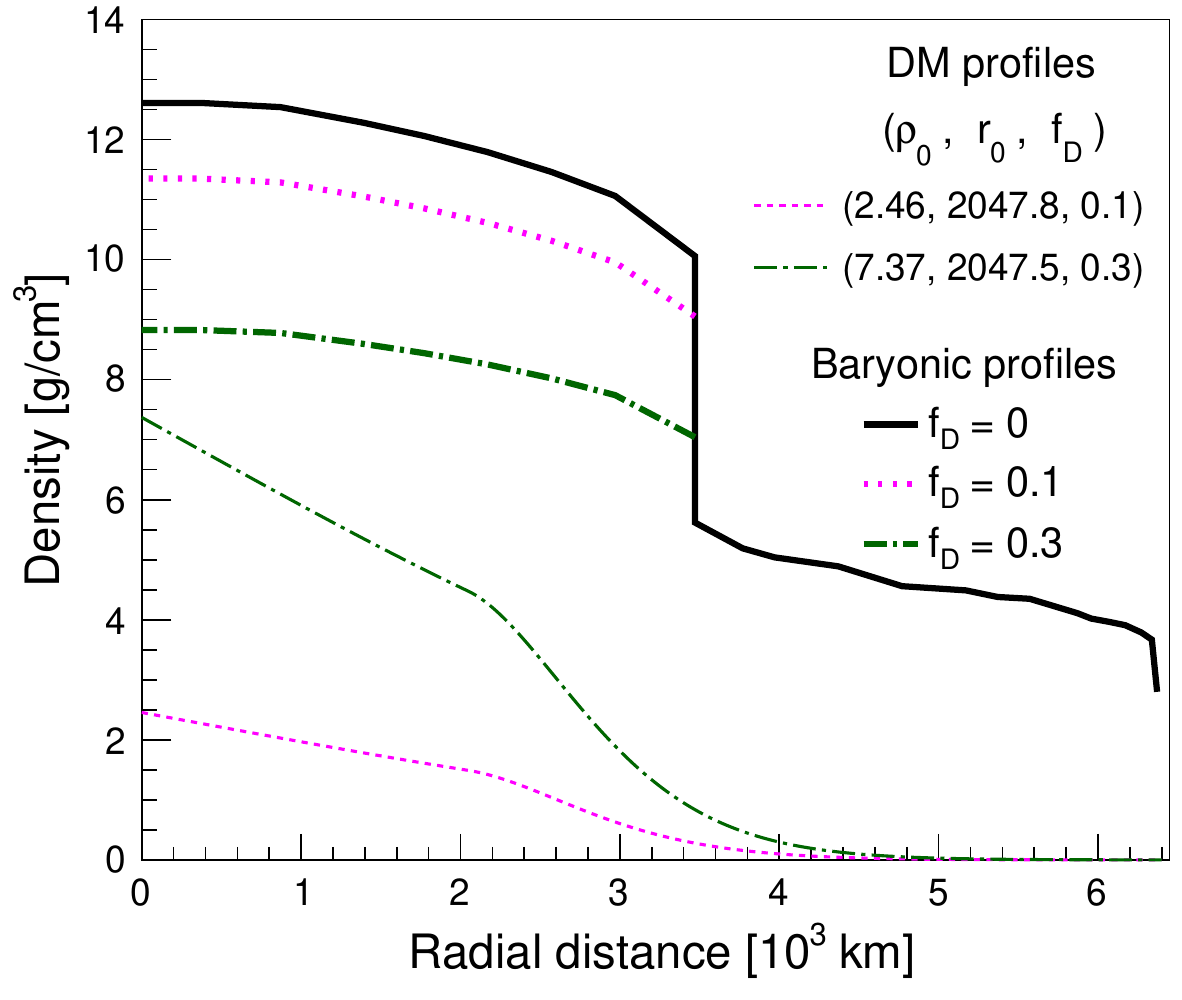}
	\caption{Representative DM profiles obeying the modified Burkert form (thin curves), and the corresponding baryonic density profiles with a constant DM mass fraction $f_\text{D} \equiv 1 - \langle f_\text{B} \rangle$ inside the core (thick curves). These representative DM profiles conserve the total mass and the moment of inertia of Earth.}
	\label{fig:DM_profile}
\end{figure*}
As noted earlier in this paper, the neutrino oscillation experiments are insensitive to the DM distribution inside Earth, as long as the total mass of DM inside Earth is given. Here, we investigate whether it is possible to have a realistic DM density profile which keep the mass of Earth $M_\text{E}$ and the moment of inertia of Earth $I_\text{E}$ invariant. We choose the DM profiles of the ``modified Burkert'' form~\cite{Burkert:1999es,Liu:2005}
\begin{equation}
	\Fontix
	\rho_\text{DM}(r)=
	\begin{cases}
	\rho_0\left[\left(1+\frac{r}{R_s}\right)\left(1+ \frac{r^2}{R_s^2} \right)\right]^{-1} & \text{if } r \leq r_0  \vspace{1mm}\\
	\rho_\text{DM}(r_0) \left(\frac{r}{r_0}\right)^\alpha \exp\left[-\frac{(r-r_0)}{r_\text{decay}}\right] & \text{if } r > r_0.  
	\end{cases}
\end{equation}
Here 
\begin{equation}
	\alpha=\frac{r_0}{r_\text{decay}}-\frac{2c^2}{(1+c^2)}-\frac{c}{(1+c)}\,, \quad \text{with} \quad c\equiv \frac{r_0}{R_s} \; ,
\end{equation}
which maintains the continuity and smoothness of the DM profile. This form has four parameters, $\rho_0$, $R_s$, $r_0$, and $r_\text{decay}$, describing the distribution of DM in the core as well as outside. Since the values of $M_\text{E}$ and $I_\text{E}$ provide two constraints, it would typically always be possible to satisfy these by choosing appropriate values of the above four parameters.  For illustration, we choose $R_s = 5000$ km and $r_\text{decay} = 0.05 R_s$, and show some representative profiles in Fig.~\ref{fig:DM_profile}, where the DM mass accounts for the decrease in the baryonic mass of the core: 
\begin{equation}
	\int_0^{R_\text{Earth}} 4 \pi r^2 \rho_\text{DM}(r) dr = \left(1 - \langle f_\text{B} \rangle \right) \cdot M_{\rm core}  \; .
\end{equation}
Thus, it is possible to have a DM distribution that contribute a given $f_\text{D} = 1 - \langle f_\text{B} \rangle$, at the same time ensuring that the total moment of inertia and total mass of Earth remains conserved.



%
\end{document}